\documentclass[prb,twocolumn,showpacs,nofootinbib]{revtex4}
\usepackage[dvips]{graphicx}
\usepackage{color,array,dcolumn}
\usepackage{amsmath}
\usepackage{amssymb}

\begin{document}

\author{P.A. Marchetti}
\affiliation{Dipartimento di Fisica ``G. Galilei'', INFN, I-35131 Padova, Italy}
\author{ G. Orso}
\affiliation{International School for Advanced Studies (SISSA), Via Beirut, 34014
Trieste, Italy}
\author{Z.B. Su}
\author{ L. Yu}
\affiliation{Institute of Theoretical Physics and Interdisciplinary Center of Theoretical
Studies,\\
Chinese Academy of Sciences, 100080 Beijing, China}
\title{IN-PLANE CONDUCTIVITY ANISOTROPY IN UNDERDOPED CUPRATES IN THE
SPIN-CHARGE GAUGE APPROACH}
\date{\today }

\begin{abstract}
Applying the recently developed spin-charge gauge theory for the
pseudogap phase in cuprates, we propose a self-consistent
explanation of several peculiar features of the far-infrared
in-plane AC conductivity, including a broad peak as a function of
frequency and significant anisotropy at low temperatures, along
with a similar temperature-dependent in-plane anisotropy of DC
conductivity in lightly doped cuprates. The anisotropy of the
metal-insulator crossover scale is considered to be responsible
for these phenomena. The obtained results are in good agreement
with experiments. An explicit proposal is made to further check
the theory.
\end{abstract}

\maketitle

The normal state properties of cuprate superconductors have shown
a number of unexpected features, particularly in underdoped
regime, where the pseudogap effects are pronounced.\cite{timusk}
One of the most recent surprises is the strong \textit{in-plane}
temperature-dependent anisotropy of DC conductivity, observed in
lightly doped untwinned LSCO and YBCO crystals.\cite{ando} The
maximal anisotropy $\rho _{b}/\rho _{a}-1$ reaches 50\% for 3\%
doped LSCO which is far too big compared with the orthorhombicity
(up to 1.7\%).\cite{ando} A \textquotedblleft
natural\textquotedblright\ explanation of this unusual behavior
would be the \textquotedblleft self-organized\textquotedblright\
charge stripe structure, proposed by a number of authors,
\cite{zaanen,white}which was also suggested to be responsible for
the occurrence of superconductivity.\cite{emery} The conductivity
is indeed higher in the stripe direction (along the \textit{a}-axis).\cite%
{matsuda} Meanwhile, Dumm \textit{et al.}~\cite{basov} recently
measured the in-plane AC conductivity of untwinned 3\% and 4\%
doped LSCO crystals. In the far infrared region at high
temperatures, the data are consistent with a simple Drude model.
Below 80 K, a broad peak appears at finite frequencies ($\Omega
\sim $ 100 cm$^{-1} $) bearing a close resemblance to the peak
found in temperature-dependent DC conductivity for the same
composition; a significant \textit{a-b}
resistivity anisotropy is observed in complete analogy with the DC case.~%
\cite{ando} These AC data seem also to support the presence of
 charge stripes.

However, there are several substantial difficulties in the intuitive
\textquotedblleft rivers of charge\textquotedblright\ interpretation: The
mean-field theory predicts the statically charge-ordered state is an
insulator,\cite{zaanen} while experimentally these lightly doped cuprates
show metallic behavior at high temperatures. Also, the observed anisotropy
ratio is too small compared with quasi-one-dimensional (Q1D) conductors,
usually showing order-of-magnitude bigger conductivity in the chain
direction. To avoid these difficulties, the \textquotedblleft electronic
liquid crystal\textquotedblright\ scenario of meandering stripes\cite%
{emery,zaanen99} is invoked to induce metallic conductivity and to reduce
the expected anisotropy. This does not solve the problem, either. In fact, a
closer examination of the data\cite{ando,basov} reveals that the anisotropy
effect is most pronounced in the limit $\omega ,T\rightarrow 0$, in
contradiction with the fluctuating stripe picture: One would anticipate much
bigger effect of anisotropy at some characteristic frequencies of the stripe
fluctuations, rather than the static limit. This was also pointed out in
Ref. \onlinecite{basov}. Also, as evident from the experimental curves,\cite%
{ando,basov} the major source of the in-plane conductivity
anisotropy is due to the shift of the Metal-Insulator Crossover
(MIC) scale, and a stronger anisotropy is found in the
\textit{localized}, instead of \textit{metallic}
regime. Up to now, many authors attribute this MIC to the disorder effect,%
\cite{basov94} or more specifically to localization in Q1D systems.\cite%
{basov} However, there is a fundamental difficulty in that
approach: There is \textit{only one mobility edge} in
disorder-induced localization for anisotropic systems \cite{rice},
at least within the scaling theory. This means the system cannot
be localized in one direction, while delocalized in another.

Recently, we have developed a spin-charge gauge approach to
describe the pseudogap phase in cuprate superconductors,
particularly focusing on the MIC phenomena.\cite{mar98,mar00} In
this approach\cite{mar98} based on spin-charge decomposition
applied to the 2D $t$-$J$ model, the spinon dynamics is described
by a non-linear $\sigma$-model with a theoretically derived mass
gap $m_s \sim J (\delta |\ln \delta|)^{1/2}$, where $J$ is the
exchange integral, $\delta$ the doping concentration; the holon is
fermionic with ``small'' Fermi surfaces $(\epsilon_F \sim t
\delta)$ (with $t$ as the hopping integral) centered around $(\pm
\pi/2, \pm \pi/2)$ in the Brillouin zone and a ``Fermi-arc''
behavior for the spectral weight. Both holons and spinons are
strongly scattered by gauge fluctuations. As an effect of gauge
interaction, the spinon mass picks up a dissipative term: $m_s
\rightarrow M_{T}=(m_s^2- i c {T/ \chi})^{1/2}$, where $\chi \sim
t \delta^{-1}$ is the diamagnetic susceptibility and $c$ a
numerical constant. This shift in turn introduces a dissipation in
the spinon-gauge sector, whose behavior dominates the low energy
physics of the system. The competition between the mass gap and
the dissipation is responsible for the MIC, giving rise to a broad
peak in the DC conductivity. At low temperatures the
antiferromagnetic (AF) correlation length $\xi \sim m_s^{-1}$ is
the determining scale of the problem, leading to localizing
behavior, while at higher temperatures, the de Broglie wave length
$\lambda_T \sim (\chi/T)^{1/2}$ becomes comparable, or even
shorter than $\xi$, giving rise to metallic conductivity. Hence in
this approach the MIC is mainly due to \textit{correlation and AF
order}, rather than \textit{disorder} effect.

In this Report we generalize this approach to frequency-dependent
phenomena and show the AC-conductivity exhibits a maximum as
function of frequency in an exact analogy with the DC conductivity
maximum due to MIC. This outcome is fully understandable: In the
presence of an external electromagnetic
field of frequency $\Omega $, that frequency will replace the temperature $T$%
, playing the role of cutoff parameter, for $\Omega >T$ at low
temperatures. Moreover, if we assume the AF correlation length
$\xi $ is anisotropic, we argue that as a consequence the same is
true for the MIC  scale, then the in-plane conductivity anisotropy
as well as the close parallel between the above two sets of
experimental data can be interpreted in a self-consistent way.

Let us now analyze the theory in more details. In a gauge approach the
physical resistivity is calculated using the Ioffe-Larkin addition rule:\cite%
{Iof} $\rho= \rho_s + \rho_h$, where $\rho_s$ and $\rho_h$ are the spinon
and holon resistivities, respectively. In order to evaluate the relevant
current-current correlation functions, one first integrates over holons and
spinons, and finds that the gauge propagator in the scaling limit, $%
\omega,q,\omega/q \rightarrow 0$, has a Reizer singularity\cite{Reizer} for
the transverse component $\langle A^T A^T \rangle (\omega,\vec q) \sim (-
\chi |\vec q |^2 + i \kappa {\omega / |\vec q|})^{-1} $ where $\kappa \sim
\delta$ is the Landau damping.

The effect of Reizer singularity on gapless fermions is
subdominant; it has been analyzed in Ref. \onlinecite{Lee} and at
finite $T$ it gives a scattering rate for the holon of order
$T^{4/3}$ instead of the usual Fermi liquid result $\sim T^2$. To
include, non-perturbatively, the effect of gauge fluctuations in
the spinon current-current correlation functions, we expand the
spinon propagator in Feynman paths, as justified by the mass gap
and we integrate over velocity fields in the eikonal approximation.\cite%
{mar00} Being gauge invariant, the correlator of the spinon current depends
only on the gauge field strength. In the scaling limit only the magnetic
components $F_{ij}$ are relevant (see Ref.~\onlinecite{mar00}); the
corresponding propagator at finite $T$ is given by
\begin{eqnarray}  \label{prop}
&&\langle F_{ij}(x) F_{rs}(0) \rangle = [\delta_{ir}
\delta_{js}-\delta_{is}\delta_{jr}] \int \frac{d\omega}{2\pi}  \notag \\
&&\cdot \int \frac{d\vec{k}}{(2\pi)^2} \, \frac{|\vec{k}|^{2} e^{ i \vec{k}%
\cdot\vec{x}-i \omega x^{0} }}{i\frac{\omega} {|\vec{k}|} \kappa -\chi |\vec{%
k}|^{2}} \coth (\frac{\omega}{2T}),
\end{eqnarray}
where $x=(\vec{x},x^0)$ with $x^0$ as the time variable. In the presence of
an external electric field (the probe in linear response theory) with
frequency $\Omega$, the integration over $\omega$ should be cutoff at $%
|\omega|\leqq \Lambda=$max($T,\Omega)$. (We further assume, for technical
reasons, that $\Lambda x^0 \ll 1$, as justified \textit{a posteriori}, see
Ref.~\onlinecite{mar00}).

Let us first focus on the static response and set $\Lambda=T$. As $\omega <T$%
, we approximate $\coth \frac{\omega}{2T}\simeq \frac{2T}{\omega}$, then the
$\omega$-integration in (\ref{prop}) becomes
\begin{equation}  \label{dc}
\int_{0}^{T} \frac{d\omega}{2 \pi}\frac {2T}{(\omega)^2+(\frac{\chi {|{\vec k%
}|}^3}{\kappa})^2},
\end{equation}
eventually leading to an evaluation of the large-scale
``magnetic'' propagator: $-i (T/4\pi\chi) Q_{T}^{2}
\exp{(-Q_{T}^{2} {|\vec{x}|}^{2}/4)}$, where
$Q_T=({T\kappa}/{\chi})^{1/3}$ is the inverse of an anomalous skin
depth, identified as the length-scale of gauge fluctuations. The
main effect of the gauge interaction on the spinon propagator is a
renormalization of the mass term in the exponent in the limit $x_0
\gg |\vec x|$:
\begin{equation}  \label{om}
m_s x^0 \rightarrow \sqrt{m_s^2-\frac{ T}{\chi}f( \frac{|{\vec x}|Q_T}{2}) }%
x_0 -\frac{ T}{2\chi} Q_T^2 g(\frac{|{\vec x}|Q_T}{2} ) \frac{x_0^2}{m_s^2},
\end{equation}
where $f$ and $g$ are regular functions, whose explicit integral
representations are given in Ref. \onlinecite{mar00}. In the evaluation of
the spatial Fourier transform of the current-current correlator, the shift (%
\ref{om}) eventually leads to a complex saddle point for $|\vec x| Q_T$ with
absolute value $C \sim O(1)$ and phase factor  $e^{i \pi/4}$, which in turn
introduces a dissipation term in the spinon gap:
\begin{equation}  \label{shift}
m_s \rightarrow M_T=(m_s^2- i c {T / \chi})^{1/2} ,
\end{equation}
where $ic=f(Ce^{i \pi/4})$. The competition between the gap term $m_s^2$ and
the dissipation $T/\chi$ leads to a MIC upon the decrease of temperature,
yielding a broad peak in the DC conductivity for $T \sim \chi m_s^2 \sim
(t/\delta) |\delta \ln \delta| \sim t |\ln \delta|$, thus shifting to lower
temperature upon doping increase. More precisely the behavior derived for
the DC conductivity is given by:\cite{mar00}
\begin{equation}  \label{dcco}
\sigma(T) \sim {\left (\frac{\delta}{f^{\prime\prime}(Ce^{i \pi/4}) |M_T|}%
\right)}^{1/2} \sin(\frac{1}{2}\arg{M_T }),
\end{equation}
where $f^{\prime\prime}$ means the second derivative.

\begin{figure}[tbp]
\includegraphics[width=4cm,angle=270]{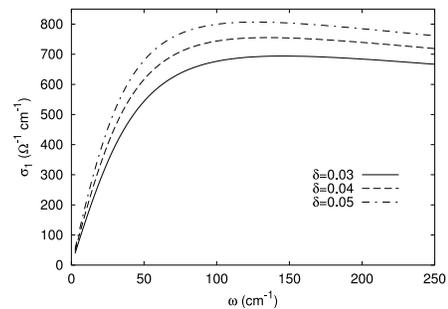}
\caption{Theoretically calculated frequency dependence of the AC
conductivity for different dopings: $\protect\delta=0.03$ (full line), $%
\protect\delta=0.04$ (dashed) and $\protect\delta=0.05$ (dotted).}
\label{fig-1}
\end{figure}

\begin{figure}[tbp]
\includegraphics[width=4cm,angle=270]{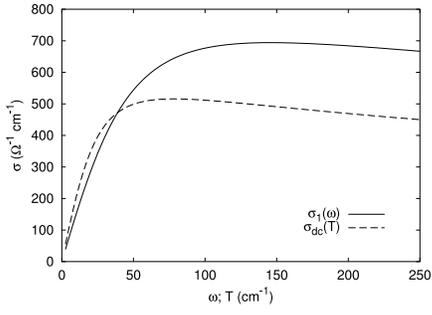}
\caption{Calculated frequency dependence of the AC conductivity for $\protect%
\delta=0.03$. Also shown is the corresponding DC conductivity as a function
of temperature (in cm$^{-1}$).}
\label{fig-2}
\end{figure}

We turn now to the AC conductivity at $T=0$ and set $\Lambda=\Omega$. In
this case $\coth \frac{\omega}{2T}$ is replaced by sgn${\omega}$ and the $%
\omega$-integration in (\ref{prop}) becomes:
\begin{equation}  \label{oac}
\int_{0}^{{\Omega}} \frac{d\omega}{2 \pi}\frac {\omega}{(\omega)^2+(\frac{%
\chi {|\vec k|}^3}{\kappa})^2}.
\end{equation}
Up to logarithmic accuracy, one finds for the magnetic propagator
at large scales:  $-i (T/4\pi\chi) Q_{\Omega}^{2}\lambda
\exp{(-Q_{\Omega}^{2} {|\vec{x}|}^{2}/4)}$
where $Q_{\Omega}=({\Omega\kappa}/{\chi})^{1/3}$ and $0 < \lambda
\lesssim 1/2$, as follows from comparing (\ref{oac}) and
(\ref{dc}). Repeating the steps of the DC calculations with this
parameter $\lambda$ included, we find as the analog of
(\ref{shift}):
\begin{equation}  \label{shift0}
m_s \rightarrow M_{\Omega}=(m_s^2- i c \lambda {\Omega / \chi})^{1/2}.
\end{equation}
For $\Omega \ll 2m_s$, one easily obtains for the AC conductivity:
\begin{equation}  \label{sac}
\sigma_{1}(\Omega) \sim {\left(\frac{\delta}{\lambda
f^{\prime\prime}(Ce^{i \pi/4}) |M_{\Omega}|}\right)}^{1/2}
\sin(\frac{1}{2}\arg{M_{\Omega} }).
\end{equation}
We see that the behavior of the AC conductivity at $T=0$ is rather similar
to that of the DC conductivity, with a broad peak corresponding to a MIC,
hardening and shifting to lower temperature upon doping increase, (see Fig.~%
\ref{fig-1}). Although the replacement of $T$ by $\Omega$ as
cutoff yields {\it a priori} only an order estimation, the
presence of the factor $\lambda$ (mainly coming from the factor
$\coth{\frac{\omega}{2T}}$) suggests that the position of the peak
in the AC conductivity is shifted by a factor $\approx
\lambda^{-1}$ to higher frequencies with respect to the DC case
and its value is enhanced by a factor $\approx
\lambda^{-\frac{1}{2}}$
( see Fig.~\ref{fig-2}), roughly in agreement with experimental data.\cite%
{basov}

The finite temperature behavior of the dynamical conductivity is as follows:
for $T\ll \Omega $ we get only a small correction to the damping term in (~%
\ref{shift0}) of order $(T/\Omega )^{5/3}$ while for $\Omega
\lesssim T$ essentially the $\Omega =0$ result applies and the
conductivity will be frequency independent and equal to the DC
value. Therefore upon temperature increase, the MIC peak is
expected to shrink asymmetrically and eventually disappears from
the spectrum, a behaviour consistent with  experiments
\cite{basov}. The limits of validity of the approximations
involved in the
calculation of the spinon correlation functions are $m_{s}^{2}\geq c{\frac{%
\Lambda }{\chi }}\geq m_{s}Q_{\Lambda }$, where the lower bound
comes from the effectiveness of the saddle point at $|\vec{x}|\sim
Q_{\Lambda }^{-1}$. When expressed in terms of temperature, this
yields a range between a few tens and few hundreds of Kelvin. We
expect that the upper limit corresponds to a crossover to a new
\textquotedblleft strange metal\textquotedblright\ phase, analyzed
in \cite{MOSY}, where the  $\pi$-flux lattice is melt and the
``metallic'', linear in $T$ resistivity is recovered. In some
sense the pseudogap phase is on the ``insulating'' side of the
MIC, and the description adopted here is  a rather good
approximation near the MIC, but the ``high temperature
asymptotics'' $\sim T^{1/4}$ is not correct, although it
reproduces at lower temperature the inflection point in
resistivity found experimentally. The above calculations do not
take into account the holon contribution to the physical
conductivity, but that is of order $\Lambda ^{4/3}$, hence
negligible for small cutoff $\Lambda $.

\begin{figure}[tbp]
\includegraphics[width=4cm,angle=270]{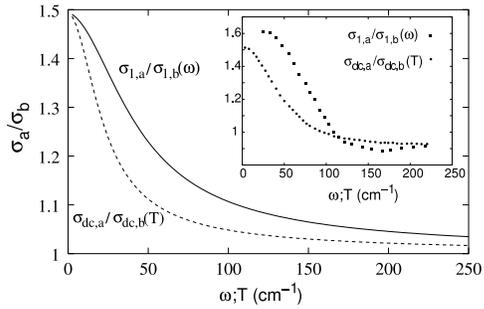}
\caption{Calculated frequency dependence of the AC conductivity
anisotropy ratio for $\protect\delta=0.03$. The corresponding DC
ratio as a function of temperature (in cm$^{-1}$) is shown with
dashed line. Inset shows the corresponding DC and AC data, taken
from Refs.\protect \onlinecite{ando} and \protect
\onlinecite{basov}, respectively. } \label{fig-3}
\end{figure}

Let us finally discuss how the in-plane resistivity anisotropy found~\cite%
{ando,basov} in untwinned single crystals of La$_{2-x}$Sr$_{x}$CuO$_{4}$ ($%
x=0.02-0.04$) can fit into our scheme. The neutron scattering experiments
have revealed incommensurate magnetic structure in lightly doped LSCO
samples ($\delta \leq 0.05$).\cite{wakimoto} Unlike the superconducting LSCO
compounds where the deviation of the elastic magnetic scattering peaks from $%
(\pi ,\pi )$ is along the $a,b$ directions in the tetragonal basis,\cite%
{yamada} these peaks are rotated by 45 degrees around $(\pi ,\pi )$, \textit{%
i.e.,} they are located along the $b^{\ast }$ axis in the
orthorhombic basis. Moreover, from the half-width of the
scattering peaks one can determine the magnetic correlation length
in different directions. As a big surprise,
one finds the correlation length strongly anisotropic. In particular, for $%
\delta =0.024,\;\xi _{a^{\ast }}^{\prime }=94.9A,\;\xi _{b^{\ast
}}^{\prime }=39.9A$.\cite{matsuda} The authors of Ref.
\onlinecite{matsuda} interpreted this result as due to stripe
formation along $a$-axis, but no quantitative argument was given.
This behavior is fully consistent with the magnetic susceptibility
anisotropy, observed in untwinned lightly doped LSCO crystals (up
tp 3\% doping).\cite{lavrov} We do not have a quantitative
microscopic theory to consider the anisotropy of the AF
correlation length $\xi $, but we can see how such anisotropy can
be included in our scheme and explore its consequences. Suppose
the hole distribution is anisotropic (which may come from the
underlying stripe structure), say the average distance between
holes is bigger along the $a$ axis, so does the distribution of
vortices on the AF background.  To use the nonlinear-$\sigma$
model treatment of spinons,  we can rescale the spatial
coordinates. The result will be almost the same as in the
isotropic case,\cite{mar00} except for a coefficient $\alpha$ in
the spinon dispersion
$\sqrt{m_s^2+\alpha^2v_s^2k_x^2+v_s^2k_y^2}$,  which reflects the
ratio of the AF correlation lengths in different directions. Since
the spinon mass $m_{s}$ is inversely proportional to the
correlation length, we can effectively interpret this as
$m_{s,a}<m_{s,b}$. To calculate the anisotropic conductivity we
need to modify the entire scheme. However,  the major effect can
be grasped without detailed calculation. The diamagnetic
susceptibility $\chi$ and Landau damping $\kappa$ due to holons
will change, but only very slightly, since they come from angular
integration. On the other hand, in the saddle point calculation of
the path integral the effect is more pronounced, so the values of
integrals $f$, $g$  and hence the numerical factor $c$, mentioned
above, will also change. In our spin-charge gauge approach the
combination $c \chi m_{s}^{2}={\frac{t}{6\pi }}|\ln \delta |r$ is
crucial in determining the MIC scale. If we assume that the basic
results of our theory developed for the 2D isotropic model survive
the generalization to anisotropic case as outlined above, one
would anticipate the parameter $r$ to be also anisotropic. In view
of the AF correlation length anisotropy we expect $r_{a}/r_{b}<1$.
As a consequence, the peak
in $\sigma _{a}$ will be shifted to lower temperature with respect to $%
\sigma _{b}$, as follows from Eq.(\ref{shift}); the anisotropy ratio $\sigma
_{a}/\sigma _{b}$ will show a sharp increase near the MIC and saturates as $%
T\rightarrow 0$, in agreement with experiments.~\cite{ando} The
same phenomenon occurs for the AC conductivity at low
temperatures, where the factor $\lambda $ makes the anisotropy
ratio even bigger. To estimate this enhancement, we extract the
ratio $r_{a}/r_{b}$ by fitting the DC data (the extracted value
$r_{a}/r_{b}=0.725$). We can then use (\ref{sac}) to evaluate the
corresponding anisotropy ratio for AC conductivity without
introducing any additional parameters; a comparison with the
experimental curve is shown in Fig.~\ref{fig-3}. This anisotropy
is less pronounced than the experimentally observed value for AF
correlation length, as quoted above for $\delta =0.024$, but this is
consistent with the above scheme, where we expect that part of the anisotropy effect in the
combination $c \chi m_{s}^{2}$ has already been cancelled by other
effects.

It is true that these peculiar features in the in-plane
conductivity are ``due to modifications of the dynamics of the
metallic carriers, and not due to the opening of a charge
gap'',\cite{basov} since there is no charge gap in doped Mott
insulators.  What we have shown here is that these modifications
are due to the presence of a gap in the spin  excitations and its
competition with the dissipation  which is different from the
disorder induced localization. For the same reason, the statement
of \textit{ only one mobility edge} in the scaling theory,\cite{rice}
 does not apply here. As we learned from the authors of Ref.  %
\onlinecite{basov},\cite{basov1} the anomalous behavior of AC conductivity
is observed only up to 6\% doping. It is understandable that the anisotropy
due to stripes is not present since they are rotated by 45 degrees beyond
6\% doping, and their orientation is alternating between $a$ and $b$
directions in adjacent layers. However, the disappearance of the
low-frequency peak cannot be explained by the stripe interpretation. On the
contrary, this is very natural in our interpretation since the MIC is not
observed in samples beyond 6\% doping in the absence of magnetic field. Now
we make an explicit proposal: To do the AC experiment in the presence of a
magnetic field which would suppress superconductivity and reveal MIC. If the
deviation from the Drude behavior reappears, that would be a confirmation of
our interpretation.

To conclude we have shown that the peculiar in-plane anisotropy of
DC and AC conductivity observed in the lightly doped cuprates can
be explained in a unified, self-consistent manner within the
spin-charge gauge approach, and the key ingredient is to attribute
the MIC to the correlation effect. The anisotropy of the AF
correlation length, and consequently the MIC scale provides a
rather natural explanation of the observed conductivity
anisotropy, being most pronounced in the limit $\Omega, T
\rightarrow 0$ which is very difficult to explain based only on
the stripe existence. In fact, this is a crucial experiment to
distinguish the disorder- and correlation-induced MIC.

We would like to sincerely thank Y. Ando and D. Basov for sending us the
paper (Ref. \onlinecite{basov}) prior to publication and very helpful
correspondence.

\clearpage \clearpage

\end{document}